\documentclass[twoside,twocolumn,9pt]{article}
\usepackage{lmodern}
\usepackage{newunicodechar}
\newunicodechar{ }{\,}
\usepackage{extsizes}
\usepackage[normalem]{ulem} % restores standard \underbar behavior 
\usepackage[super,sort&compress,comma]{natbib} 
\usepackage[version=3]{mhchem}
\usepackage[left=1.5cm, right=1.5cm, top=1.785cm, bottom=2.0cm]{geometry}
\usepackage{balance}
\usepackage{mathptmx}
\usepackage{amsfonts}
\usepackage{sectsty}
\usepackage{amssymb}
\usepackage{pifont}
\usepackage{graphicx} 
\usepackage{lastpage}
\usepackage[format=plain,justification=justified,singlelinecheck=false,font={stretch=1.125,small,sf},labelfont=bf,labelsep=space]{caption}
\usepackage{subcaption}
\usepackage{booktabs}
\usepackage[utf8]{inputenc}
\usepackage{float}
\usepackage{fancyhdr}
\usepackage{fnpos}
\usepackage[english]{babel}
\usepackage{textcomp}
\usepackage{newunicodechar}

\usepackage{array}
\usepackage{droidsans}
\usepackage{charter}
\usepackage[T1]{fontenc}
\usepackage[dvipsnames]{xcolor}
\usepackage{setspace}
\usepackage[compact]{titlesec}
\usepackage{hyperref}

\definecolor{cream}{RGB}{222,217,201}

\begin{document}

%%%PAGE STYLE%%%
\pagestyle{fancy}
\thispagestyle{plain}
\fancypagestyle{plain}{
\renewcommand{\headrulewidth}{0pt}
}

%%%FONT SIZES%%%
\makeatletter
\renewcommand\LARGE{\@setfontsize\LARGE{15pt}{17}}
\renewcommand\Large{\@setfontsize\Large{12pt}{14}}
\renewcommand\large{\@setfontsize\large{10pt}{12}}
\renewcommand\footnotesize{\@setfontsize\footnotesize{7pt}{10}}
\makeatother

%%%FOOTNOTE FORMATTING%%%
\makeFNbottom
\renewcommand{\thefootnote}{\fnsymbol{footnote}}
\renewcommand\footnoterule{\vspace*{1pt}% 
\color{cream}\hrule width 3.5in height 0.4pt \color{black}\vspace*{5pt}}
\makeatletter 
\renewcommand\@biblabel[1]{#1}            
\renewcommand\@makefntext[1]{\noindent\makebox[0pt][r]{\@thefnmark\,}#1}
\makeatother 

%%%SECTION FORMATTING%%%
\setcounter{secnumdepth}{5}
\sectionfont{\sffamily\Large}
\subsectionfont{\normalsize}
\subsubsectionfont{\bf}
\titlespacing*{\section}{0pt}{4pt}{4pt}
\titlespacing*{\subsection}{0pt}{15pt}{1pt}

%%%FIGURE AND SPACING%%%
\renewcommand{\figurename}{\small{Fig.}~}
\setstretch{1.125}
\setlength{\skip\footins}{0.8cm}
\setlength{\footnotesep}{0.25cm}
\setlength{\jot}{10pt}
%%%END OF PAGE SETUP%%%

%%%FOOTER%%%
\fancyfoot{}
%\fancyfoot[LO,RE]{\vspace{-7.1pt}\includegraphics[height=9pt]{head_foot/LF}} % Commented out for arXiv submission
\fancyfoot[RO]{\footnotesize\sffamily\thepage}
\fancyfoot[LE]{\footnotesize\sffamily\thepage}
\fancyhead{}
\renewcommand{\headrulewidth}{0pt} 
\renewcommand{\footrulewidth}{0pt}

%%%LAYOUT SPACING%%%
\setlength{\arrayrulewidth}{1pt}
\setlength{\columnsep}{6.5mm}
\setlength\bibsep{1pt}

%%%FIGURE RULES%%%
\makeatletter 
\newlength{\figrulesep} 
\setlength{\figrulesep}{0.5\textfloatsep} 
\newcommand{\topfigrule}{\vspace*{-1pt}%
\noindent{\color{cream}\rule[-\figrulesep]{\columnwidth}{1.5pt}}}
\newcommand{\botfigrule}{\vspace*{-2pt}%
\noindent{\color{cream}\rule[\figrulesep]{\columnwidth}{1.5pt}}}
\newcommand{\dblfigrule}{\vspace*{-1pt}%
\noindent{\color{cream}\rule[-\figrulesep]{\textwidth}{1.5pt}}}
\makeatother
%%%END OF FIGURE RULES%%%

%%%TITLE, AUTHORS AND ABSTRACT%%%
\twocolumn[
\begin{@twocolumnfalse}
\sffamily
\begin{center}
\vspace{0.5cm}
\noindent\LARGE{\textbf{Quantum Synthetic Data Generation for Industrial Bioprocess Monitoring}}
\vspace{0.5cm}

\noindent\large{Shawn M. Gibford\footnotemark[1], Mohammad Reza Boskabadi\footnotemark[1], Christopher J. Savoie\footnotemark[2], and Seyed Soheil Mansouri\footnotemark[1]\footnotemark[3]}

\vspace{0.4cm}
\begin{minipage}{0.9\textwidth}
\normalsize
Data scarcity and sparsity in bio-manufacturing poses challenges for accurate model development, process monitoring, and optimization. We aim to replicate and capture the complex dynamics of industrial bioprocesses by proposing the use of a Quantum Wasserstein Generative Adversarial Network with Gradient Penalty (QWGAN-GP) to generate synthetic time series data for industrially relevant processes. The generator within our GAN is comprised of a Parameterized Quantum Circuit (PQC). This methodology offers potential advantages in process monitoring, modeling, forecasting, and optimization, enabling more efficient bioprocess management by reducing the dependence on scarce experimental data. Our results demonstrate acceptable performance in capturing the temporal dynamics of real bioprocess data. We focus on Optical Density, a key measurement for Dry Biomass estimation. The data generated showed high fidelity to the actual historical experimental data. This intersection of quantum computing and machine learning has opened new frontiers in data analysis and generation, particularly in computationally intensive fields, for use cases such as increasing prediction accuracy for soft sensor design or for use in predictive control.
\end{minipage}
\end{center}
\vspace{0.6cm}
\end{@twocolumnfalse}
]
%%%END OF TITLE, AUTHORS AND ABSTRACT%%%

%%%FONT SETUP%%%
\renewcommand*\rmdefault{bch}\normalfont\upshape
\rmfamily

\vspace{-1cm}

%%%AUTHOR AFFILIATIONS%%%
\footnotetext[1]{\textit{Department of Chemical and Biochemical Engineering, Technical University of Denmark, Søltofts Plads, Building 228A, 2800 Kongens Lyngby, Denmark}}
\footnotetext[2]{\textit{SiC Systems Inc., Franklin, TN, USA}}
\footnotetext[3]{\textit{Corresponding Author Email: seso@dtu.dk}}

%%%MAIN TEXT%%%%

\section{Introduction}
Machine Learning (ML) and Artificial Intelligence (AI) have emerged as transformative tools in diverse scientific disciplines, from cancer research to financial analysis. \cite{clinical_iqbal_2021} \cite{vakili2024quantumcomputingenhancedalgorithmunveils} \cite{artificial_ahmed_2022} \cite{Orus2019QuantumCF} Despite this widespread adoption, complex systems, such as manufacturing, drug discovery, and formulated products, \cite{emerson2021} have lagged notably in fully embracing ML/AI methodologies compared to other fields that have led to tools such as ChatGPT (OpenAI), \cite{openai2024gpt4technicalreport} Gemini (Google), \cite{geminiteam2025geminifamilyhighlycapable} and Claude (Anthropic) \cite{claude42025}  that are currently being used as generalized intelligence assistants. This discrepancy is not due to the complete absence of ML applications, as evidenced by several review articles, \cite{khanal2023artificial, singh2022role, narayanan2020bioprocessing} but rather reflects the unique challenges posed by the complex and non-linear process where the acquisition of high volume data is challenging, including high dimensionality, inherent variability, and complex underlying chemical, physical, or biological mechanisms. \cite{boskabadi2025virtual}

\begin{figure*}
    \centering
    \includegraphics[height = 7cm]{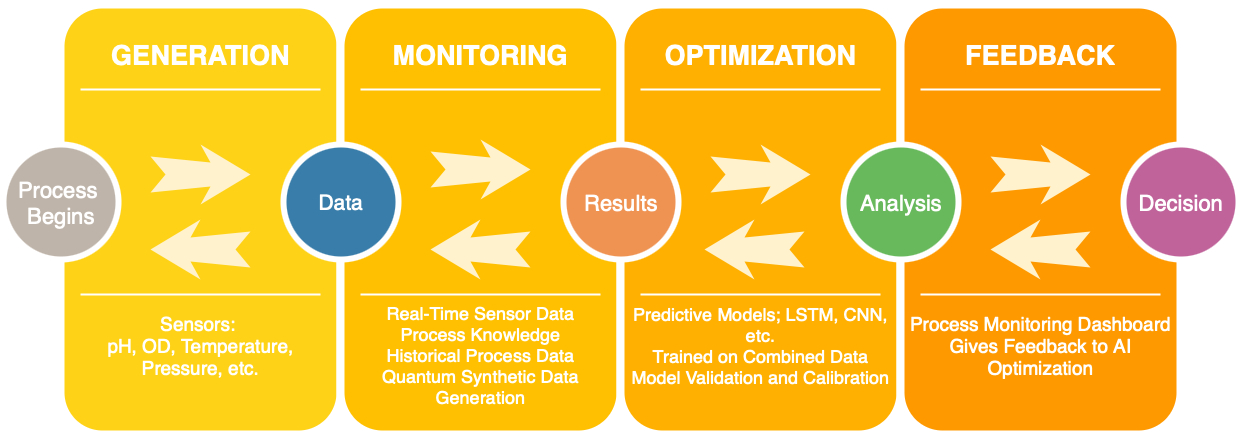}
    \caption{Conceptual diagram of how we envision using quantum generative AI within the industrial bioprocess}
    \label{fig:concept_diagram}
\end{figure*}

One example of a complex system with data availability challenges is bioprocess engineering, a field crucial for the production of biofuels, pharmaceuticals, and sustainable materials. Data analytics and machine learning applications to this field often grapples with limited data sets due to the time-consuming and resource-intensive nature of bioprocess experiments, \cite{stoneOptimize} and the lack of offline / online measurement technologies available. This scarcity of data can hinder the development of accurate predictive models, and the problem is compounded by the absence of measurement of hard-to-quantify process quality indicator variables (QIVs), which are essential for process optimization and real-time monitoring.  In this field, data are usually generated in two ways, with physical sensor(s) to measure an observable state, or a model that describes the states (e.g. a soft sensor). The objective is to monitor the process and improve productivity with one or a combination of these two approaches.

Critical QIVs in bioprocesses include biomass concentration, which is often estimated through optical density measurements or viable cell counts but requires complex calibration and suffers from interference in high-density cultures. \cite{kiviharju2008biomass} Product concentration and quality, such as monoclonal antibody titers and glycosylation patterns in mammalian cell culture, remain challenging to measure online and typically require offline analytical methods such as High-Performance Liquid Chromatography
(HPLC) or mass spectrometry.  Metabolic activity indicators, including specific growth rates, substrate consumption rates, and metabolic flux distributions, are fundamental QIVs that are difficult to quantify in real-time without sophisticated analytical platforms. Furthermore, process performance metrics such as volumetric productivity, product yield coefficients, and process consistency indicators are complex QIVs that integrate multiple process variables and require advanced data analytics for accurate estimation. Environmental stress indicators, including osmolality changes, shear stress effects, and pH fluctuations' impact on cell viability, represent another class of QIVs that significantly influence process outcomes but are challenging to monitor continuously. 

The integration of ML with bioprocess engineering has been gradually expanding, with applications spanning process monitoring, \cite{peng2025machinelearningmethodssmall} optimization, \cite{sharma2025ai} and control. \cite{mondal2023review} These approaches have demonstrated potential to address the limitations of traditional mechanistic models, particularly when dealing with the non-linear dynamics and multivariate nature of biological systems. The field continues to face significant barriers to a wider implementation, including data quality issues, interpretability concerns, and the need for domain-specific adaptations of existing ML frameworks.

A fundamental challenge limiting the widespread adoption of machine learning in bioprocess engineering is data scarcity. \cite{lim2023opportunities, duongtrung2023when} Unlike fields such as computer vision or natural language processing, where datasets may be comprised of millions of samples, bioprocess datasets are typically constrained by several factors. Bioprocess experiments, particularly on the pilot and production scales, involve substantial costs in terms of equipment, raw materials, personnel, and time. \cite{shariatifar2025digital} \cite{exploring_hernndezromero_2025} A single bioreactor at pilot or production scale run may take weeks to complete and can be costly, often on the order of several hundreds of thousands of dollars or more, severely limiting the number of experimental conditions that can be feasibly explored, e.g., scale-up operations. \cite{exploring_hernndezromero_2025} Biological systems exhibit inherent stochasticity and sensitivity to environmental conditions, resulting in significant batch-to-batch variability even under nominally identical operating conditions.  This variability introduces noise into the data sets and requires additional replications to establish statistical confidence, further increasing data scarcity. Often, bio-manufacturing to produce food or pharmaceutical products operates under stringent regulatory frameworks that can limit experimental flexibility and the implementation of exploratory approaches that generate diverse datasets.  The need to maintain product quality and process consistency often results in conservative operational strategies that produce limited data variability. Commercial bioprocesses represent significant intellectual property, which results in limited data sharing across the industry.  While fields such as computer vision benefit from open datasets with millions of examples, bioprocess data typically remains siloed within individual companies or research groups. These factors collectively result in datasets that are often too small to effectively train complex ML models, particularly deep learning architectures that typically require substantial amounts of diverse training data. The problem is further compounded for rare event detection and process fault diagnosis, where example data may be extremely limited or entirely absent. 

Synthetic data generation is a promising approach to address data scarcity challenges in bioprocess engineering. \cite{liu2024bestpracticeslessonslearned} By creating artificial data sets that faithfully represent the statistical properties and dynamic behaviors of real bioprocesses, synthetic data approaches can potentially overcome limitations in data availability while avoiding confidentiality concerns. Several approaches to the generation of bioprocess data have been explored and listed in Table \ref{tbl:bioprocess_methods}. Note, however, that, to the best of the authors' knowledge, we have not been able to identify evidence of application of quantum generative AI or quantum enhanced synthetic data generation in bioprocess engineering.

First-principles models have been used to simulate bioprocess dynamics under various conditions, generating synthetic time series data.\cite{vidossich2021role} Although this approach ensures physical consistency, its fidelity is limited by the accuracy of the underlying model and its ability to capture process variability and uncertainty. Methods such as Gaussian process regression and multivariate statistical models can generate synthetic data based on estimated probability distributions.  \cite{chokwitthaya2020applying} These approaches excel at reproducing statistical properties, but often struggle to capture complex temporal dependencies and rare events.

Machine learning models such as variational autoencoders (VAEs) and generative adversarial networks (GANs) can learn directly from the available data to generate synthetic samples. \cite{wang2018esrganenhancedsuperresolutiongenerative, akkem2024comprehensive} These approaches offer flexibility in generating diverse datasets without requiring detailed mechanistic understanding, although their performance is highly dependent on the quality and quantity of training data. Combining mechanistic knowledge with data-driven techniques represents a promising direction for the generation of synthetic bioprocess data.  

\begin{table*}
\small
  \caption{\ Approaches for Bioprocess Synthetic Data Generation}
  \label{tbl:bioprocess_methods}
  \begin{tabular*}{\textwidth}{@{\extracolsep{\fill}}p{3.5cm}p{5cm}p{4cm}p{3.5cm}}
    \hline
    \textbf{Approach} & \textbf{Description} & \textbf{Advantages} & \textbf{Limitations}\\
    \hline
    Mechanistic Model-Based \& 
First-principles models &
simulate bioprocess dynamics under various conditions \cite{vidossich2021role, banga2025mechanistic}& 
Ensures physical consistency & 
Limited by model accuracy, understanding of pathways/interactions, and ability to capture process variability \\
\midrule
   Statistical Approaches & 
Gaussian Process regression and multivariate statistical models generate data based on estimated probability distributions \cite{kim2023, miletic2025utility} & 
Excel at reproducing statistical properties & 
Struggle with complex temporal dependencies and rare events \\
\midrule
    Data-Driven Approaches & 
Machine learning models (VAEs, GANs) learn directly from available data \cite{goodfellow2014generativeadversarialnetworks, akkem2024comprehensive} & 
Flexible generation without requiring mechanistic understanding & 
Performance depends heavily on training data quality and quantity\\
\midrule
Hybrid Approaches & 
Combine mechanistic knowledge with data-driven techniques \
\cite{albino2024hybridmodel} & 
Incorporates physical constraints and biological relationships & 
Complexity in implementation and model integration \\
    \hline
  \end{tabular*}
\end{table*}

The utility of synthetic data extends beyond simply augmenting training datasets. In bioprocess development, synthetic data generation enables in silico exploration of process design spaces, risk assessment, development of monitoring strategies, and operator training without consuming physical resources. For established processes, synthetic data can facilitate the development and validation of monitoring algorithms, including fault detection systems for rare but critical process deviations. \cite{liu2024bestpracticeslessonslearned}

Despite these advances, significant challenges remain in the generation of high-quality data. Accurate capture of complex temporal dependencies, incorporation of mechanistic knowledge constraints, ensuring physical realizability, and validating the quality of synthetic data, for example. Addressing these challenges requires novel methodological approaches that can effectively combine the strengths of both classical and quantum computational paradigms.

With the advent of quantum technologies, Quantum Generative Adversarial Networks  (QGANs) have become powerful tools to generate synthetic data in various domains. \cite{Dallaire_Demers_2018, Lloyd_2018, Zoufal_2019}  QGANs leverage the inherent properties of quantum systems to provide potential advantages over classical generative models. The exponential state space of quantum systems enables a more efficient representation of complex probability distributions, particularly those with multimodal or high-dimensional characteristics. \cite{zoufal2021generativequantummachinelearning} Quantum superposition and entanglement allow QGANs to explore multiple solution paths simultaneously during training, potentially leading to faster convergence and reduced susceptibility to mode collapse. \cite{romero2019variationalquantumgeneratorsgenerative, Amin_2018} Additionally, quantum interference effects can enhance the model's ability to capture subtle correlations and nonlinear relationships present in real-world data, making them particularly suited for complex temporal dependencies found in bioprocess systems. \cite{Liu_2019, Stamatopoulos_2020} Furthermore, quantum algorithms demonstrate theoretical advantages in certain optimization landscapes, suggesting that QGANs may overcome some of the training instabilities that plague classical GANs when working with limited datasets. \cite{farhi2014quantumapproximateoptimizationalgorithm, Cerezo_2021}.

Rudolph et al. \cite{rudolph2022generationhighresolutionhandwrittendigits} demonstrated the ability of QGANs to generate images of higher quality from the MNIST dataset than its classical counterpart. In \cite{orlandi2024enhancing}, quantum-generated financial data was shown to improve the performance of predictive models compared to using historical data alone.   However, classical GANs may struggle to capture the intricate temporal dependencies and non-linear dynamics inherent in complex time series data. \cite{goodfellow2014generativeadversarialnetworks} QGANs, using the principles of quantum computation, offer a promising alternative to address these limitations. \cite{lloyd2013quantumalgorithmssupervisedunsupervised}

The principal contributions of this work are fourfold: 

\begin{itemize}
    \item Closed-Loop Decision-Driven Pipeline. A novel eight-stage workflow that integrates sensor feasibility checks, mechanistic models, data-driven learning, and quantum synthetic data generation in a single feedback loop.

    \item  QGAN for Process Monitoring. A QGAN architecture optimized for the generation of high-fidelity synthetic samples in biochemical process applications directly improves the prediction performance of downstream estimators.

    \item Empirical Validation on Industrial Bioprocesses. Quantitative demonstrations on real-world cultivation datasets show that this approach is successful for generating synthetic dateset, simulating the real world dynamic of data.

    \item Open Science and Reproducibility. The complete code, trained models, and datasets have been released at github.com/shawngibford/qgan for transparency, independent verification, and acceleration of subsequent advances in quantum‐enhanced chemical and biochemical engineering.
\end{itemize}

The remainder of this paper is structured as follows. Section~\ref{sec:Classical and Quantum Machine Learning} reviews relevant concepts in classical and quantum machine learning, highlighting their roles in synthetic data generation. Section~\ref{sec:Framework} details our quantum-enhanced framework for generating synthetic bioprocess data. Section~\ref{sec:Experimental Procedure} describes the experimental setup and outlines the novel aspects of our approach. In Section~\ref{sec:Results}, we present and evaluate the performance of the generated data. Section~\ref{sec:Discussion} discusses the implications of our findings for bioprocess monitoring and control, summarizes our conclusions, and proposes directions for future research.

\section{Classical and Quantum Machine Learning}
\label{sec:Classical and Quantum Machine Learning}
The increasing complexity of industrial bioprocess engineering demands precise monitoring and optimization, yet data scarcity remains a significant bottleneck. Traditional bioprocesses, particularly on pilot scale operations, generate limited datasets due to high operational costs, time constraints, and regulatory requirements. This data shortage impedes the development of robust process models and optimization strategies.

Although classical synthetic data generation techniques offer partial solutions, they often struggle to capture the intricate temporal dependencies and non-linear relationships characteristic of bioprocess systems.\cite{goodfellow2014generativeadversarialnetworks} Recent advances in quantum computing, particularly in generative modeling, present promising opportunities to better capture these correlations and dependencies. QGANs have shown superior performance in the generation of high-quality synthetic data \cite{esteban2017realvaluedmedicaltimeseries} for healthcare applications, notably in cancer treatment research. \cite{vakili2024quantumcomputingenhancedalgorithmunveils} These successes suggest similar potential for bioprocess engineering.

Time-series analysis represents a critical component of bioprocess monitoring and optimization, and ML approaches demonstrate considerable success in handling the temporal dependencies inherent in bioprocess data. Long-Short-Term Memory (LSTM) networks, a specialized form of recurrent neural networks, have proven particularly effective for bioprocess time series forecasting due to their ability to capture long-term dependencies while mitigating the problem of vanishing gradients. \cite{beck2024xlstmextendedlongshortterm} These networks utilize memory cells and gating mechanisms to selectively retain or discard information across time steps, making them well-suited for bioprocess dynamics prediction.

Autoencoders, comprising an encoder and decoding components, have been used to reduce dimensionality, extract features, and detect anomalies in bioprocess data. \cite{bank2021autoencoders} By learning compressed representations of input data, autoencoders facilitate the identification of latent patterns and the reconstruction of high-dimensional bioprocess measurements. Their unsupervised nature makes them particularly valuable when labeled data is scarce, a common scenario in bioprocess development.

GANs represent another significant advancement in ML applicable to bioprocess engineering. Consisting of generator and discriminator networks engaged in a competitive training process, GANs have demonstrated remarkable capabilities in the generation of synthetic data. \cite{goodfellow2014generativeadversarialnetworks, wang2018esrganenhancedsuperresolutiongenerative} In bioprocess contexts, GANs offer potential solutions to data scarcity issues by generating realistic bioprocess profiles that preserve temporal correlations and process dynamics. Recent adaptations, such as TimeGAN \cite{yoon2019TimeGAN} and C-RNN-GAN \cite{mogren2016crnngancontinuousrecurrentneural}, have further refined GAN architectures for time-series applications, enhancing their utility for bioprocess simulation.

For bioprocess synthetic data generation, QGANs may address several limitations of classical approaches. The enhanced representational capacity of quantum circuits could enable more faithful modeling of complex multimodal distributions characteristic of bioprocess data. \cite{Zoufal_2019} Likewise, quantum interference effects might allow QGANs to better capture subtle correlations in limited datasets, potentially improving the quality of synthetic data even with the data scarcity typical of bioprocess applications \cite{he2025qganbaseddataaugmentationhybrid}.

However, classical GANs face significant challenges when applied to bioprocess data. Training instability, mode collapse, and difficulty in capturing complex multimodal distributions can limit their effectiveness in generating high-fidelity bioprocess synthetic data.  Additionally, the relatively small size of the available bioprocess data sets often leads to overfitting or insufficient knowledge of the underlying data distribution. These limitations motivate the exploration of alternative computational paradigms, including quantum approaches, which may offer advantages for specific aspects of synthetic data generation.

\subsection{Classical Machine Learning}
\label{Classical ML}
Classical machine learning approaches for synthetic data generation rely on learning statistical patterns and relationships within existing datasets to produce new, realistic samples. These methods have found extensive application in bioprocess engineering, where they address data scarcity challenges through various algorithmic strategies. \cite{khanal2023artificial, khuat2023applicationsmachinelearningbiopharmaceutical}

Among these classical techniques, GANs have become a powerful tool for synthetic data generation. GANs are capable of producing highly realistic data by simultaneously training a generator and a discriminator in a competitive framework. In the following section, we provide a brief discussion of GANs, highlighting their relevance, capabilities, and limitations in the context of synthetic data generation for complex systems such as industrial bioprocesses.

\subsubsection{Generative Adversarial Networks (GANs)} 
GANs represent the foundation of modern synthetic data generation. \cite{goodfellow2014generativeadversarialnetworks} The GAN framework consists of two neural networks engaged in a competitive minimax game: a generator $G$ that learns to produce synthetic data, and a discriminator $D$ that distinguishes between real and synthetic samples. The objective function is formulated as:

\begin{equation}
\min_G \max_D V(D,G) = \mathbb{E}_{x \sim p_{\text{data}}(x)}[\log D(x)] + \mathbb{E}_{z \sim p_z(z)}[\log(1-D(G(z)))]
\end{equation}

where $p_{\text{data}}(x)$ represents the real data distribution and $p_z(z)$ is the noise distribution. For bioprocess applications, specialized architectures such as TimeGAN \cite{yoon2019TimeGAN} incorporate recurrent components to better capture temporal dependencies inherent in process data.

\begin{figure}
    \centering
    \includegraphics[width=1\linewidth]{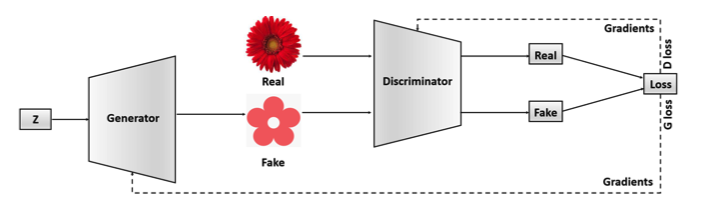}
    \caption{Illustration of classical GAN architecture}
    \label{fig:classical_gan}
\end{figure}

Classical approaches have demonstrated success in various bioprocess applications, including fermentation profile generation, sensor data augmentation, and fault scenario simulation.  However, their effectiveness is constrained by training stability issues, limited ability to capture multimodal distributions, and sensitivity to hyperparameter selection. \cite{Arjovsky2017, gulrajani2017improved} These limitations become particularly pronounced when working with the small, high-dimensional datasets typical of bioprocess engineering, motivating exploration of quantum-enhanced alternatives. 

\subsubsection{Wasserstein GAN with Gradient Penalty}
\label{WGANGP}
The advent of GANs by Goodfellow et al. \cite{goodfellow2014generativeadversarialnetworks} marked a paradigm shift in generative modeling, introducing a novel framework wherein two neural networks, a generator and a discriminator, are trained in an adversarial manner. Since then, this foundational work has catalyzed a burgeoning field of research, leading to a plethora of architectural and theoretical advancements.  Despite their success in generating high-fidelity samples, the original GAN formulation is notoriously difficult to train, often suffering from issues such as mode collapse and vanishing gradients. \cite{Arjovsky2017} 

To address these training instabilities, the Wasserstein GAN (WGAN) was introduced, which uses the Earth Mover's distance to provide a more meaningful and stable loss metric. \cite{arjovsky17a} However, the initial WGAN implementation relied on weight clipping to enforce the Lipschitz constraint on the critic, a practice that was found to introduce pathological behavior and limit model capacity. A significant refinement to this approach came with the introduction of the Wasserstein GAN with Gradient Penalty (WGAN-GP). \cite{gulrajani2017improved} Instead of clipping weights, WGAN-GP imposes a soft constraint on the norm of the critic's gradient with respect to its input, Equation 2 is the mathematical formation of the objective function:

\begin{equation}
\min_G \max_D \left[ \mathbb{E}_{x \sim \mathbb{P}_r}[D(x)] - \mathbb{E}_{z \sim \mathbb{P}_z}[D(G(z))] - \lambda \mathbb{E}_{\hat{x} \sim \mathbb{P}_{\hat{x}}}[(\|\nabla_{\hat{x}} D(\hat{x})\|_2 - 1)^2] \right]
\end{equation}
The WGAN-GP framework employs a critic network $D(\cdot)$ and generator network $G(\cdot)$ in a minimax optimization scheme. The objective approximates the Wasserstein distance between the real data distribution $\mathbb{P}_r$ and the distribution of generated samples produced from noise distribution $\mathbb{P}_z$. To enforce the Lipschitz constraint, a gradient penalty term is introduced with coefficient $\lambda$ (typically set to 10), which operates on interpolated samples $\hat{x} = \epsilon x + (1-\epsilon)G(z)$ where $\epsilon \sim \text{Uniform}(0,1)$. These interpolated points, distributed according to $\mathbb{P}_{\hat{x}}$, provide a differentiable path between real and generated samples, enabling stable enforcement of the 1-Lipschitz continuity requirement without weight clipping.

This solution not only stabilizes the training process across a wide range of architectures but also mitigates the issues associated with weight clipping, leading to improved sample quality and more reliable convergence. The WGAN-GP has since become a cornerstone of modern generative modeling, influencing a wide array of subsequent research and applications. 

\begin{figure}
    \centering
    \includegraphics[width=1\linewidth]{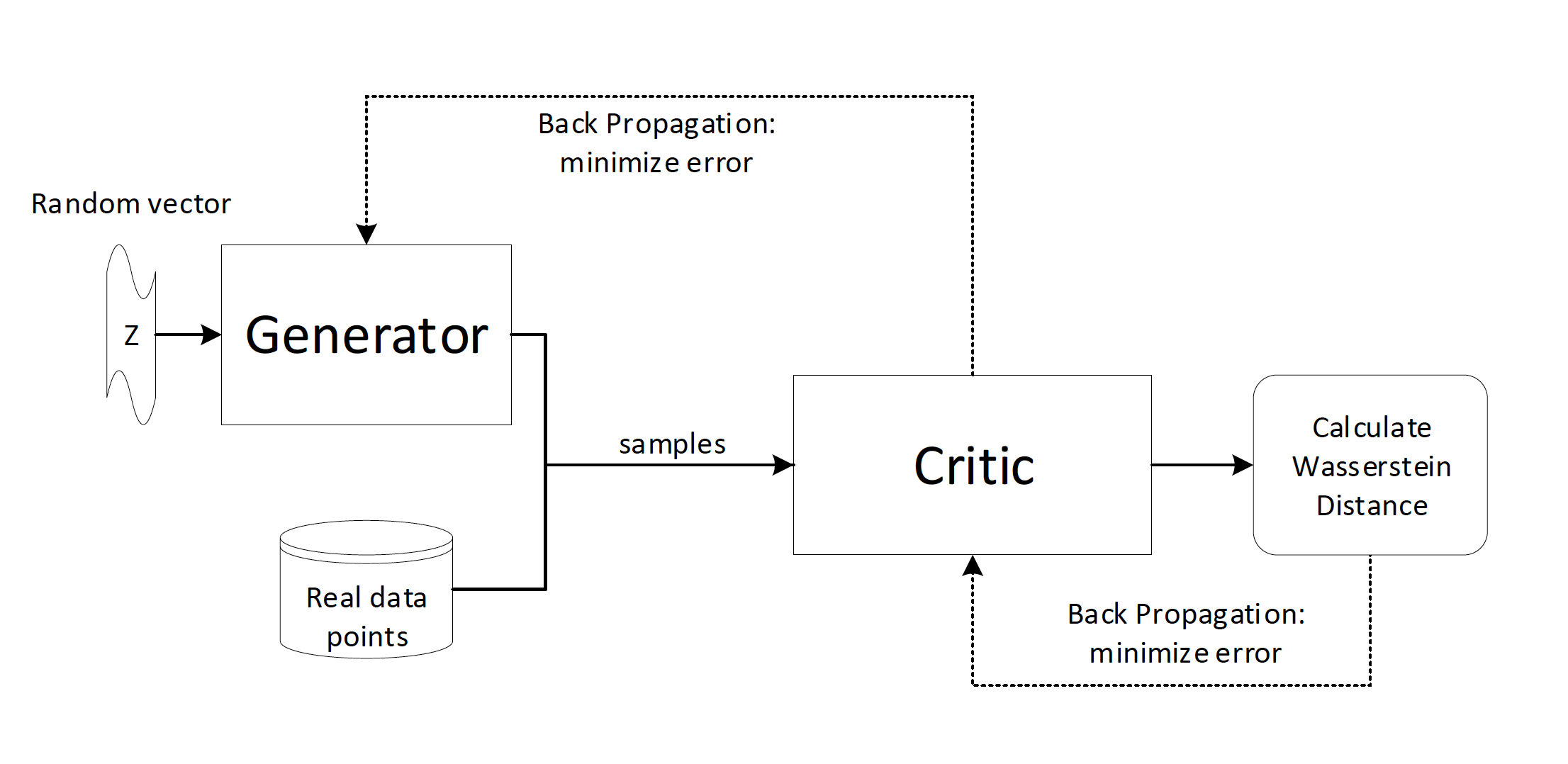}
    \caption{Classical WGAN-GP architecture}
    \label{fig:wgan}
\end{figure}

\subsubsection{Wasserstein Distance (Earth Mover's Distance)}
\label{EMD}
The Wasserstein distance, also known as the Earth Mover's Distance (EMD), provides a principled way to measure the dissimilarity between two probability distributions by quantifying the minimal cost of transforming one distribution into the other. For two probability distributions \( P \) and \( Q \) over \( \mathbb{R} \), the Wasserstein distance is defined as
\begin{equation}
W(P, Q) = \inf_{\gamma \in \Gamma(P, Q)} \int_{\mathbb{R} \times \mathbb{R}} |x - y| \, d\gamma(x,y)
\end{equation}
where \( \Gamma(P, Q) \) is the set of all joint distributions (transport plans) with marginals \( P \) and \( Q \), and \( \gamma(x,y) \) represents the amount of mass transported from \( x \) to \( y \). \cite{villani2008optimal} In the case of one-dimensional empirical distributions with cumulative distribution functions \( F_P \) and \( F_Q \), this distance simplifies to
\begin{equation}
W(P, Q) = \int_{-\infty}^{\infty} \left| F_P(t) - F_Q(t) \right| \, dt
\end{equation}

This metric captures both the magnitude and the position of differences between distributions, making it particularly robust for comparing the global properties of time series or probability distributions, as it reflects the minimal work needed to align one distribution with another.

Our QGAN implementation employs the Wasserstein loss function with gradient penalty (WGAN-GP) to ensure training stability and provide meaningful convergence metrics. The Wasserstein distance offers a principled approach to measuring distributional similarity, particularly crucial for quantum circuits where traditional loss functions may not adequately capture the quantum advantage. 

The WGAN-GP formulation is particularly well-suited for quantum generative models as it eliminates mode collapse issues that can be exacerbated by the discrete nature of quantum measurements. Additionally, the gradient penalty mechanism provides stable training dynamics essential for the hybrid quantum-classical optimization landscape, ensuring reliable convergence even when quantum circuit parameters exhibit non-convex loss surfaces. This stability is crucial for bioprocess applications, where training data is often limited and model robustness is paramount for practical deployment.

\subsection{Quantum Machine Learning}
\label{QML}
QML represents the convergence of quantum computing and machine learning, with the aim of leveraging quantum mechanical principles to improve learning algorithms. \cite{Biamonte2017} Recent advances in QML have demonstrated potential advantages in terms of expressivity, training efficiency, and generalization capabilities. \cite{kashyap2025advancesmachinelearningquantum} The field encompasses diverse approaches, including quantum-enhanced classical algorithms, quantum kernels, variational quantum algorithms, and fully quantum neural networks.

For bioprocess applications facing data scarcity challenges, QML offers several potential advantages. Quantum-enhanced sampling procedures may enable more efficient exploration of high-dimensional parameter spaces with limited data points. \cite{Cerezo_2021} Quantum kernels can potentially capture complex feature relationships in bioprocess data that classical kernels might miss. \cite{giraldo2025q2sar} Additionally, the inherent probabilistic nature of quantum systems aligns well with the need to model uncertainty in bioprocess systems with limited available data.

Quantum computing and quantum-inspired methodologies and algorithms offer potential advantages in generative modeling, with QGANs successfully applied in finance, healthcare, and optimization problems. \cite{orlandi2024enhancing} \cite{esteban2017realvaluedmedicaltimeseries} \cite{Mugel2022} Unlike classical GANs, QGANs use quantum circuits to generate probability distributions that are inherently more expressive, allowing for improved pattern generation and higher-fidelity synthetic data. \cite{rudolph2022generationhighresolutionhandwrittendigits} Despite these advantages, QGANs face challenges related to hardware limitations and scalability.

\subsubsection{Quantum GANs}
\label{qgans}
QGANs represent a hybrid quantum-classical machine learning approach where a PQC is embedded in the generative component of a GAN. First introduced by Dallaire-Demers and Killoran, QGANs leverage the unique properties of quantum computing to generate complex probability distributions more efficiently than their classical counterparts. \cite{Dallaire_Demers_2018}

A key advantage of QGANs arises from quantum superposition and entanglement, which allow quantum circuits to represent and explore high-dimensional probability spaces exponentially more compactly than classical systems. This allows QGANs to generate richer data distributions with fewer parameters compared to classical GANs. Moreover, quantum circuits inherently allow for non-linear transformations of data, which can improve the generator's ability to capture complex correlations in multi-dimensional datasets.

Recent empirical evidence suggests that QGANs can outperform classical GANs in low-data regimes or scenarios where capturing complex distributions is critical. Rudolph et al. (2022) demonstrated that QGANs can generate high-resolution images using an ion trap quantum computer, highlighting the potential of quantum models to exceed classical limitations in certain generative tasks. \cite{rudolph2022generationhighresolutionhandwrittendigits} Furthermore, QGANs have shown promise in financial modeling and synthetic data generation for time series, where accurate representations of statistical dependencies are vital. \cite{orlandi2024enhancing}

Despite these advantages, QGANs are still in the early stages of development and face several challenges. Current quantum hardware is limited by noise, decoherence, and shallow circuit depths, which constrain the complexity of trainable quantum models. Additionally, training QGANs requires hybrid quantum-classical optimization techniques, which can be computationally intensive and sensitive to initialization and learning rates. However, the theoretical potential and initial successes position QGANs as a transformative tool for the generation of synthetic data in data-scarce fields such as industrial bioprocess engineering.

QGANs extend the classical GAN framework to the quantum domain, utilizing quantum circuits for the generator or discriminator components. \cite{Lloyd_2018, Dallaire_Demers_2018} These hybrid quantum-classical models leverage quantum advantages in representing and generating complex probability distributions while maintaining the adversarial training paradigm that has proven successful in classical machine learning.

QGANs offer several potential advantages for bioprocess applications, including enhanced expressivity for representing complex distributions, potential quantum speedups during training, and the capability to generate both classical and quantum data. Recent experimental implementations have demonstrated proof-of-concept QGANs on noisy intermediate-scale quantum (NISQ) devices, \cite{rudolph2022generationhighresolutionhandwrittendigits, Huang_2021} suggesting their near-term feasibility for specialized applications.

GANs have been widely explored for time series data generation, with models such as TimeGAN \cite{yoon2019TimeGAN} and DoppelGANger \cite{Lin_2020} showing strong performance in the capture of complex temporal patterns. However, classical models are limited by computational constraints and struggle with high-dimensional probability distributions.

\section{Framework for Quantum Synthetic Data Generation}
\label{sec:Framework}

The application of the framework is demonstrated through a case study. This work is based on previous work done by Orlandi et al., \cite{orlandi2024enhancing} repurposing their model for our exploration. The quantum component of this model we used Pennylane, Xanadu's quantum SDK, for simplicity and easy of use; however, there are several others, such as IBM's Qiskit or Amazon's Bracket, that could be used. Power from our models comes from the PQC's ability to be less computationally complex, while remaining efficient. 

In the analysis of time-series data, raw values such as prices (\(P_t\)) or (\(OD_t\)), in our case, are often transformed into returns to facilitate meaningful statistical modeling. This transformation serves to normalize series of different scales and is crucial for achieving stationarity, a required property for many time series models. A more frequently used alternative is the logarithmic return given by:
\begin{equation} 
r_t = \ln(OD_t) - \ln(OD_{t-1}). 
\end{equation}
Log returns are highly favored in quantitative analysis for several reasons, most notably their convenient mathematical properties, e.g., time additivity and value compression; the sum of log returns over consecutive periods is equal to the log return over the total time span. For small changes in value, the simple and log returns are approximately equal (\(r_t \approx R_t\)), but the analytical benefits of the log returns have established them as the standard for applications. Normalization was then applied to the calculated values of OD, to a mean of zero and a standard deviation of one, bringing all values to a consistent scale using the equation:

\begin{equation} 
r_{t_{NORM}} = \frac{r_t -\mu}{\sigma}
\end{equation}
The inverse Lambert-W transform was applied to the normalized log values to adjust for the heavy tails within the distribution of the data, pushing them towards more Gaussian traits. This function is the inverse of the function $z =ue^u$, making a transformation usable for heavy-tailed datasets such as ours. Shown by Goerg, \cite{goerg2015lambert} this function allows the reshaping of a distribution to align more closely with a Gaussian distribution. Mathematically described by the equation:

\begin{equation}
    W_\delta(\nu) = sgn(\nu)\sqrt{\frac{W(\delta\nu^2)}{\delta}}
\end{equation}

where $\nu$ is a value of the dataset, $sgn(\nu)$ is the sign of $\nu$, $\delta$ is a programmable parameter, and $W$ is the Lambert W function.

Using the rolling window technique,\cite{dimoudis2023utilizing} we generated overlapping subsequences of the original data to capture temporal patterns and dependencies. This depends on two parameters: stride $(s)$ and subsequence length $(m)$, where stride is the gap between the starts of subsequences. The window length $(m)$ is set longer than $(s)$ to guarantee enough overlap. Here we have set $m$ to $10$ and $s$ to $2$, according to Orlandi et al., \cite{orlandi2024enhancing} this optimizes for a balanced overlap and computational efficiency. 

\begin{figure*}
    \centering
    \includegraphics[height = 5cm, width=1\linewidth]{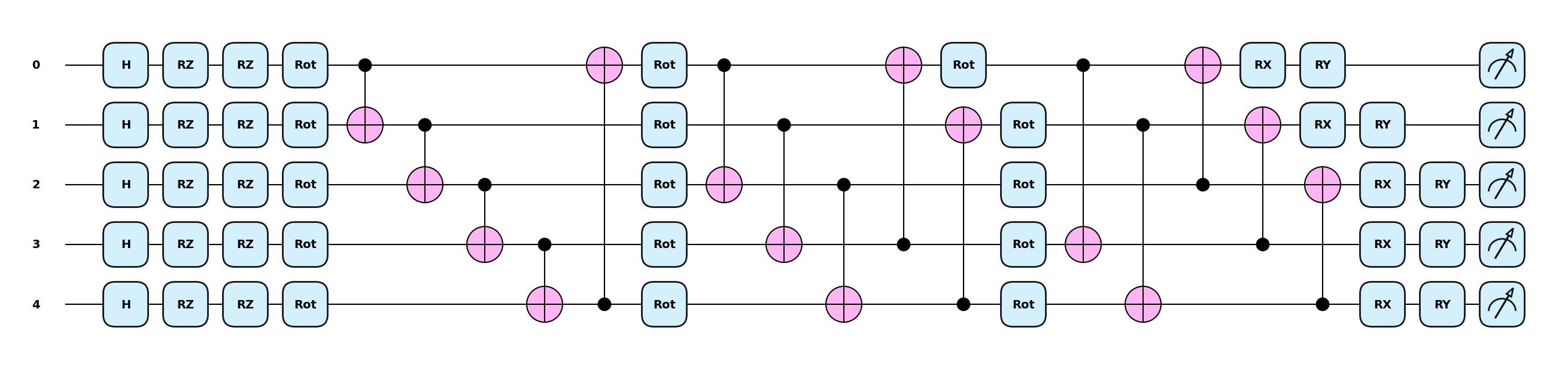}
    \caption{Quantum circuit used as generator}
    \label{fig:quantum_circuit}
\end{figure*}

It should be noted that the framework is general and may be altered however needed in other implementations in the future. What we describe is simply how we have chosen to do so in this instance. It is not necessary to use this exact set up. We trained the model for $2000$ epochs. The size of the batch, which is the number of training samples used in one iteration to update the model’s parameters, is equal to $20$. The quantum generator attempts to produce a distribution that approximates the underlying distribution. The quantum generator aims to produce a distribution that approximates the underlying distribution. The quantum circuit of our model, shown in \ref{fig:quantum_circuit}, comprises five qubits arranged in three layers. Single-qubit rotation gates are employed to control the individual qubit states, while quantum $CNOT$ gates establish entanglement between qubits. Additionally, re-uploading layers facilitate the passage of computations to the subsequent layer. This approach improves the ability of the model to capture the intricacies of our process data.

A multibasis measurement of the quantum circuit is performed using both the Pauli-X $(\sigma_x)$ and Pauli-Z $(\sigma_z)$ operators. This facilitates attaining information about the states on the $X$ and $Z$ bases, respectively. The $X$ operator looks at the superposition basis $(|+\rangle, |-\rangle)$,  while the $Z$ operator measures the quantum state in the computational basis $(|0\rangle, |1\rangle)$. Using both measurements allows a greater understanding of the behavior of the qubit. In classical computation, systems cannot be measured in two bases.

The circuit employs a dual-stage encoding strategy. An initial encoding layer sets the quantum state using random noise, where each qubit receives a unique random angle between $0$ and $2\pi$. An $R_z$ gate is applied to each qubit using these random angles as parameters to create the IQP (Instantaneous Quantum Polynomial) encoding. This randomized initialization enables quantum interference patterns and allows the generator to explore a more complex space than previous architectures used in the literature. The PQC begins with Hadamard gates applied to all qubits to create equal superposition states from the initial $|0\rangle$ states. Following initialization, a second IQP encoding layer applies parameterized $R_z$ rotations using circuit parameters, creating additional quantum polynomial interference effects that enhance the circuit's expressivity.  The circuit implements the strongly entangling layer architecture across two repeated layers. Each layer utilizes $Rot(\phi, \theta, \omega)$ gates applied to all qubits, where each $Rot$ gate is equivalent to $R_z(\phi) R_y(\theta) R_z(\omega)$, providing complete single-qubit rotational control. The entangling layers employ range-based $CNOT$ gates following the pattern $r = (layer \bmod (n_{qubits} - 1)) + 1$, where each qubit connects to its target qubit at distance $r$, creating an entanglement structure that varies between layers. Final measurement preparation applies $R_x$ and $R_y$ rotation gates to all qubits to optimize the quantum state for measurement. The circuit concludes with enhanced Pauli measurements, capturing both $X$ and $Z$ expectation values from each qubit to provide a comprehensive 10-dimensional output vector that encodes the quantum state information for classical processing. This architecture achieves 45 total parameters distributed as: 5 for IQP encoding, 30 for strongly entangled layers (15 per layer), and 10 for final measurement preparation, creating a highly expressive quantum generator with excellent statistical performance.

The Critic model is a convolutional neural network that processes sequential data to distinguish between real and generated data from the quantum generator. The network contains three convolutional layers with $64$, $128$, and $128$ filters respectively, where the increasing filter count allows extraction of more complex features and produces detailed feature maps. The first convolutional layer is designed to handle input data shaped as $(10, 1)$, where the dimension of $10$ represents the length of subsequences created through the rolling window approach, ensuring the network properly processes the temporal structure of the input data.

Each convolutional layer is followed by a LeakyReLU activation function $(\alpha = 0.1)$, which addresses the vanishing gradient issue that can cause neurons to become inactive and enables the network to learn more complex patterns through its nonlinear properties. Following the three convolutional layers, the data undergoes flattening to convert the multi-dimensional feature maps into a one-dimensional vector suitable for dense layer processing. A fully connected layer with $32$ neurons is then applied with LeakyReLU activation to capture complex data relationships. To improve generalization and reduce overfitting, a dropout layer randomly deactivates $20\%$ of neurons during training. The final output layer contains a single neuron that produces an unbounded scalar value, where higher values indicate the input resembles real data and lower values suggest generated data. This unbounded output design aligns with the use of Wasserstein distance as the cost function, see \ref{EMD}.

\begin{figure*}
    \centering
    \includegraphics[width=0.75\linewidth]{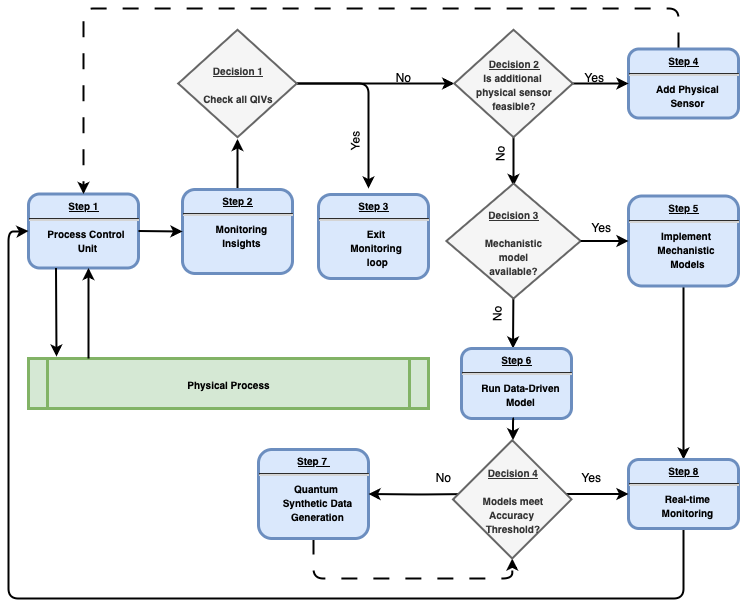}
    \caption{Decision-tree of the bioprocesses monitoring framework}\label{qgan_schemcatic}
\end{figure*}

The proposed framework is an eight-stage, decision driven pipeline that blends traditional process monitoring with quantum‐augmented data generation to produce robust soft sensors for industrial bioprocesses, Figure \ref{qgan_schemcatic} illustrates this decision tree architecture. It begins at the Process Control Unit (Step 1), where routinely measured variables—pH, temperature and dissolved oxygen, for example—are collected and forwarded to the Monitoring Insights layer (Step 2), which performs preprocessing and extracts Quality-Indicator Variables (QIVs). A first decision checkpoint verifies that all critical QIVs are present; if any are missing or unreliable, the loop returns to Step 2 for recalibration. Next, feasibility of an additional physical sensor is assessed: if viable, the sensor is installed (Step 4) and the monitoring loop is re-entered; otherwise the framework proceeds directly to the next decision. If a mechanistic model exists, it is integrated in Step 5 alongside empirical methods; if not, the framework advances straight to data-driven modeling. In Step 6, conventional soft-sensor architectures (PLS, random forests or neural networks) are trained on the combined QIVs and mechanistic outputs. A fourth decision then evaluates whether validation metrics meet predefined thresholds—if so, the soft sensor is deployed for real-time monitoring (Step 8); if not, quantum synthetic data generation is invoked (Step 7), where a variational quantum algorithm produces synthetic samples in under-represented regions of the process space. These synthetic QIVs are appended to the training set, and the enriched dataset cycles back to Step 6 until target performance is achieved. Finally, real-time QIV estimates from Step 8 are fed directly back into the Process Control Unit (Step 1), closing the loop and enabling continuous refinement of both control strategies and monitoring insights.

When monitoring proves insufficient, the framework follows a hierarchical improvement strategy. The system first considers hardware-based solutions by evaluating the feasibility of additional sensor deployment. If sensor addition is viable and cost-effective, new sensors are integrated into the monitoring system and the process returns to the initial evaluation stage.

\section{Experimental Procedure: Single-Cell Autotrophic Cultivation}
\label{sec:Experimental Procedure}
To collect time-series data the experimental work was conducted in a 20-liter photobioreactor (LUCY\textsuperscript{®}, Synoxis Algae) designed for laboratory and pilot-scale cultivation of microalgae. The system integrates SALT technology (spiral air lift tubular circulation), controlled aeration, programmable lighting, and advanced process monitoring, that are all managed by integrated automated controller. This circulation technique agitates the culture medium gently and homogenizes it effectively, which significantly reduces biofilm formation. In addition, the configuration improves gas–liquid mass transfer and ensures suitable light availability throughout the culture. The LUCY® is equipped with seamless integration of Clean-In-Place (CIP) skids. Approximately 98.5\% of the culture chamber volume is usable, with the remaining space occupied by air and CO\textsubscript{2}. The system continuously monitors and controls temperature, light intensity, pH, dissolved oxygen, optical density, gas flow rates. Data are logged automatically at defined intervals (typically every 10 minutes) by an internal data acquisition (DAQ) system for further analysis. Further details about the experimental setup and measured parameters are provided in the supplementary materials.

\begin{figure}
    \centering
    \includegraphics[width=1\linewidth]{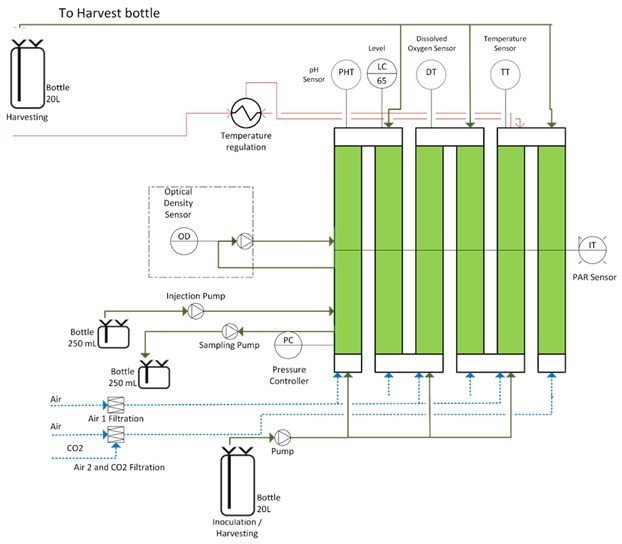}
    \caption{Schematic of the 300L LUCY \textcopyright  photobioreactor, sensors, and actuators}
    \label{fig:lucy}
\end{figure}

The LUCY\textsuperscript{®} photobioreactor setup comprises six vertical borosilicate glass tubes (6 cm OD and 120 cm length) connected in parallel via silicone lids \ref{fig:lucy}. Each tube has identical dimensions, and the modular configuration facilitates uniform illumination and mixing across the entire culture volume.

The control system uses a touchscreen controller with programmable logic to automatically monitor and adjust the cultivation conditions. It continuously measures key parameters such as pH, temperature, optical density, and pressure to track the status of the culture in real time. Feedback control loops are implemented for critical variables such as pH (controlled by $CO_2$ dosing), temperature (controlled by the thermal rod), and light intensity (regulated through programmable LED sequences). Based on these measurements, it regulates important processes including the intensity of the LED lighting, the flow of air through two separate aeration lines (AIR1 and AIR2), the addition of $CO_2$, the heating and cooling of the system, and the operation of pumps to maintain optimal growth conditions. Sensor and actuators used in the set-up are listed in Table \ref{tab:lucy_sensors}. 

\begin{table}[htbp]
\centering
\caption{\textbf{Sensors and actuators integrated into the LUCY® photobioreactor}}
\label{tab:lucy_sensors}
\resizebox{\columnwidth}{!}{%
\begin{tabular}{>{\centering\arraybackslash}p{3cm}>{\centering\arraybackslash}p{6cm}}
\toprule
\textbf{Component type}& \textbf{Description} \\
\midrule
pH sensor&
Continuous measurement of culture pH; controlled through $CO_2$ injection for pH regulation\\
\midrule
$CO_2$ injection valve& 
Supplies $CO_2$ on-demand for pH correction\\
\midrule
Temperature sensor& 
Continuous measurement of culture temperature; controlled by heater/chiller\\
\midrule
Thermal regulation rod& 
Submerged rod connected to a closed-loop water system for heating/cooling\\
\midrule
 Pressure Sensor&Monitoring of internal pressure to prevent overpressure\\
 \midrule
 Air injection valves (Air1 and Air2)&Air1 provides main aeration and $CO_2$ delivery; Air2 prevents biofilm formation\\
 \midrule
 Dissolved oxygen sensor&Continuous measurement of DO levels for assessing respiration and growth\\
 \midrule
 Optical Density (OD) sensor&In-line measurement of biomass concentration at 880 nm wavelength\\
 \midrule
 LED light panels&White-spectrum (5000 K) illumination; programmable intensity and photoperiod\\
 \midrule
 Dosing pumps (P1, P2)&Low-flow peristaltic pumps for nutrient addition and sampling (20 mL/min)\\
 \midrule
 Dosing pump (P3)&High-flow peristaltic pump for medium loading or harvest (660 mL/min)\\ 
\bottomrule
\end{tabular}%
}
\end{table}

As is depicted in \ref{fig:lucy}, Air1 injects compressed air and $CO_2$through tubes for 58 minutes of each hour. Air2 injects air through the tubes for 2 minutes in an hour, to minimize biofilm formation on reactor wall. Temperature control is achieved with a thermal regulation rod connected to a heater/chiller loop. The OD sensor operates continuously at 880 nm and, in auto mode, periodically purges and analyzes the circulating culture. output values are reported in arbitrary units based on logarithmic attenuation, with calibration performed using fresh culture medium as a blank. Dissolved oxygen measurements are captured in parallel, providing additional insight into metabolic activity and gas exchange efficiency. Lighting is provided by programmable LED panels surrounding the tubes. Light can be configured for constant output, diurnal cycles (e.g., 16/8), or stepwise ramps. Sampling and replenishment are carried out using peristaltic pumps P1 and P2 for withdrawing samples or adding nutrients, while pump P3 handles feeding or harvesting. All sampling lines are made of silicone tubing and must be chemically disinfected before and after use. Cleaning is performed automatically using programmed sequences that include rinsing, alkaline washing with Deptal MP Hypred\textcopyright (20 mL/L), acid disinfection with Deptil PA5 Hypred\textcopyright (5 mL/L), and final rinsing. The cleaning solutions circulate through all wetted parts of the system, including the pumps, tubing, and growth chamber, to remove organic and inorganic residues and ensure proper sanitization before the next cultivation cycle.

\section{Results}
\label{sec:Results}

\begin{figure*}
    \centering
    \includegraphics[width=1\linewidth]{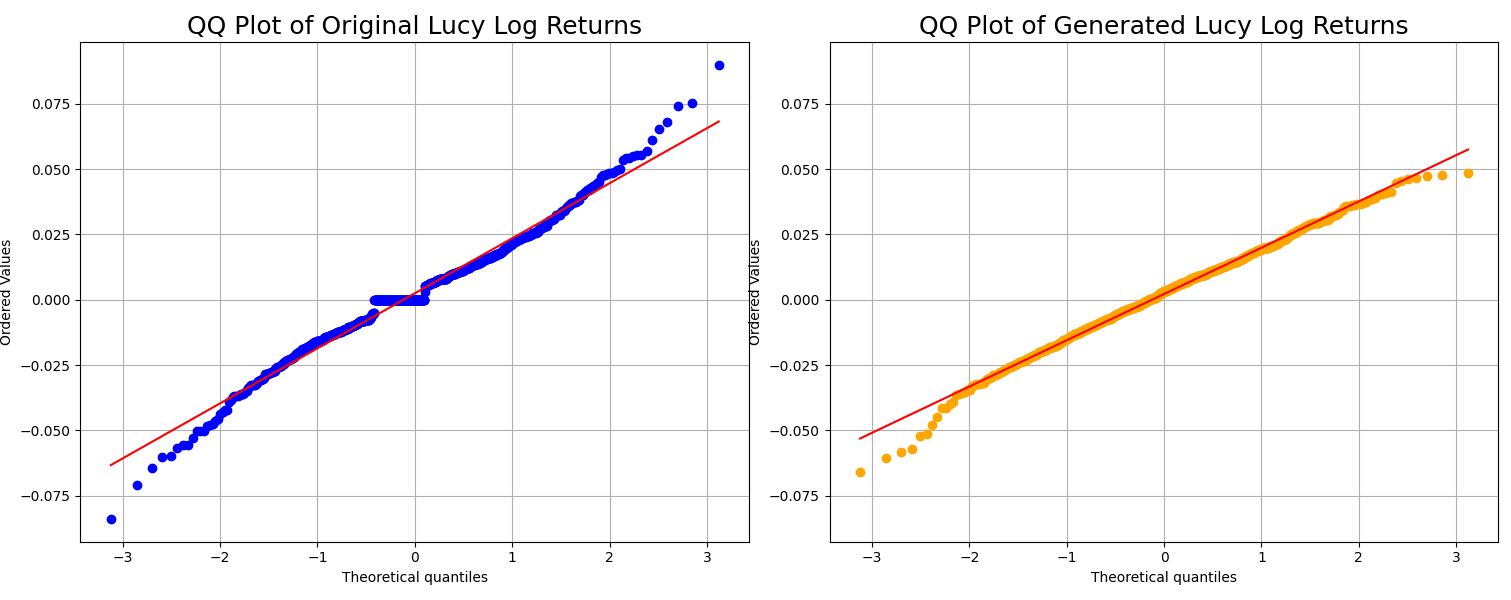}
    \caption{Quantile-Quantile (QQ) plot comparison.} 
    \label{fig:QQ}
\end{figure*}

The Quantile-Quantile (QQ) plot analysis shown in Figure~\ref{fig:QQ} reveals a strong normality alignment for both historical and synthetic data sets. The QQ plot of the original data shows an excellent adhesion to the theoretical normal distribution line in most quantiles, with minor deviations only in the extreme tails (theoretical quantiles beyond ±2). The synthetic data demonstrate similarly strong normality characteristics, maintaining close alignment with the theoretical line throughout the central region while exhibiting slightly different tail behavior. Both datasets display the characteristic S-curve pattern indicating heavier tails than a perfect normal distribution, confirming our earlier observations about non-normal distributional properties. QQ plots comparing the normality characteristics of original bioprocess data versus QGAN-generated synthetic data. Both datasets show strong adherence to normal distribution with characteristic heavy-tail deviations in extreme quantiles.

\begin{figure*}
    \centering
    \includegraphics[width=1\linewidth]{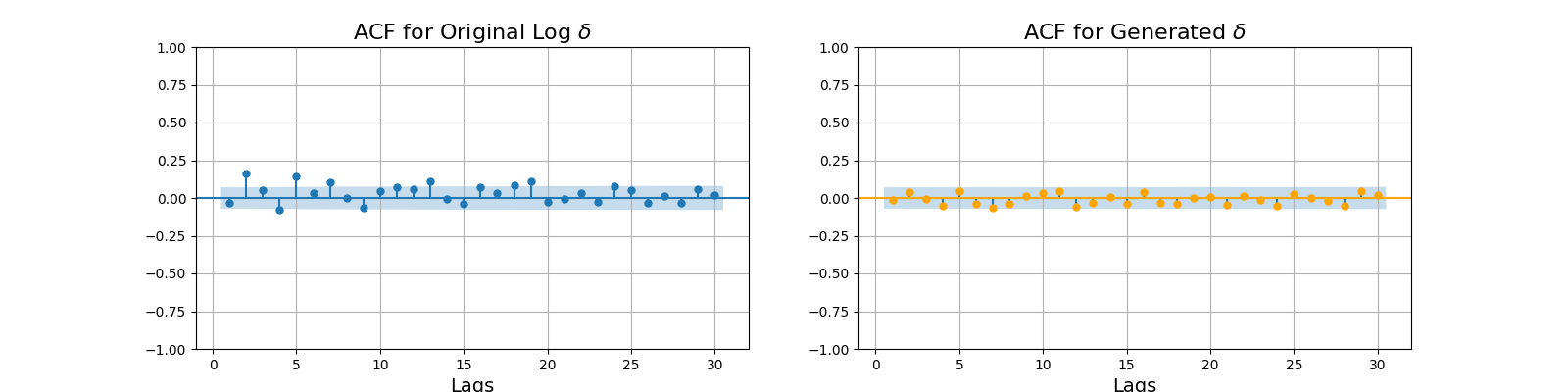}
    \caption{Auto-correlation Function (ACF) comparison.} 
    \label{fig:ACF}
\end{figure*}

\begin{figure*}
    \centering
    \begin{subfigure}[t]{0.48\textwidth}
        \centering
        \includegraphics[width=\linewidth]{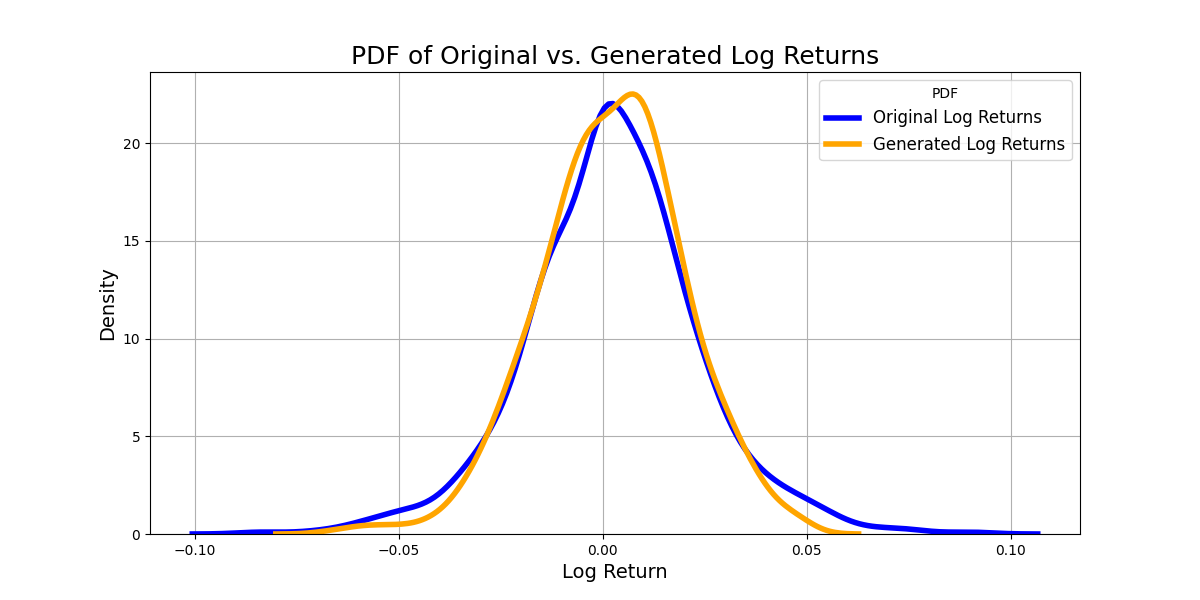}
        \caption{Probability Density Function (PDF) comparison}
        \label{fig:PDF}
    \end{subfigure}
    \hfill
    \begin{subfigure}[t]{0.48\textwidth}
        \centering
        \includegraphics[width=\linewidth]{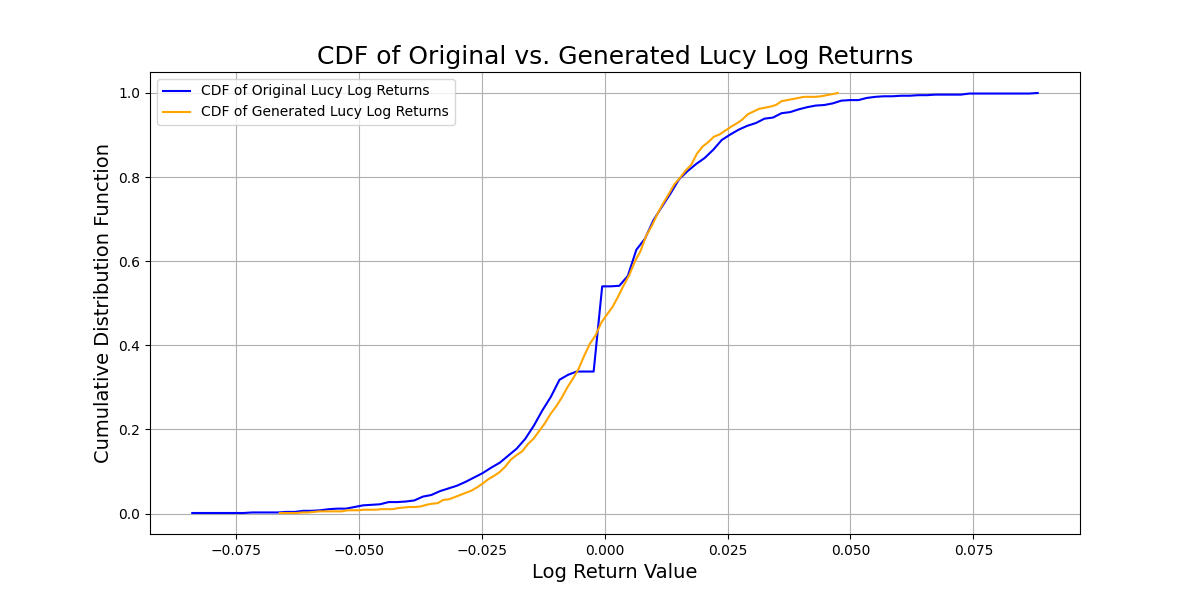}
        \caption{Cumulative Density Function (CDF) comparison}
        \label{fig:CDF}
    \end{subfigure}
    \caption{Statistical distribution comparisons}
    \label{fig:distributions}
\end{figure*}

The auto-correlation function analysis presented in Figure~\ref{fig:ACF} demonstrates the temporal dependencies preserved in our synthetic data generation process. This analysis confirms that the QGAN maintains the correlation structure inherent in the original bioprocess time series. Comparison of auto-correlation functions between original bioprocess data and QGAN-generated synthetic data, demonstrating preservation of temporal correlation structures in the synthetic dataset.
Comparisons of the probability density function and the cumulative density function shown in Figures~\ref{fig:PDF} and~\ref{fig:CDF} further demonstrate strong distributional fidelity between historical and synthetic data. Although the synthetic data exhibit slightly higher peak concentration and reduced tail variability compared to the original distribution, the overall shapes and central tendencies are well-preserved. (a) PDF comparison showing preservation of distributional characteristics with minor variations in peak concentration and tail behavior. (b) CDF analysis demonstrating excellent alignment across most quantile ranges, with deviations occurring primarily in extreme value regions.

\begin{figure}
    \centering
    \includegraphics[width=1\linewidth]{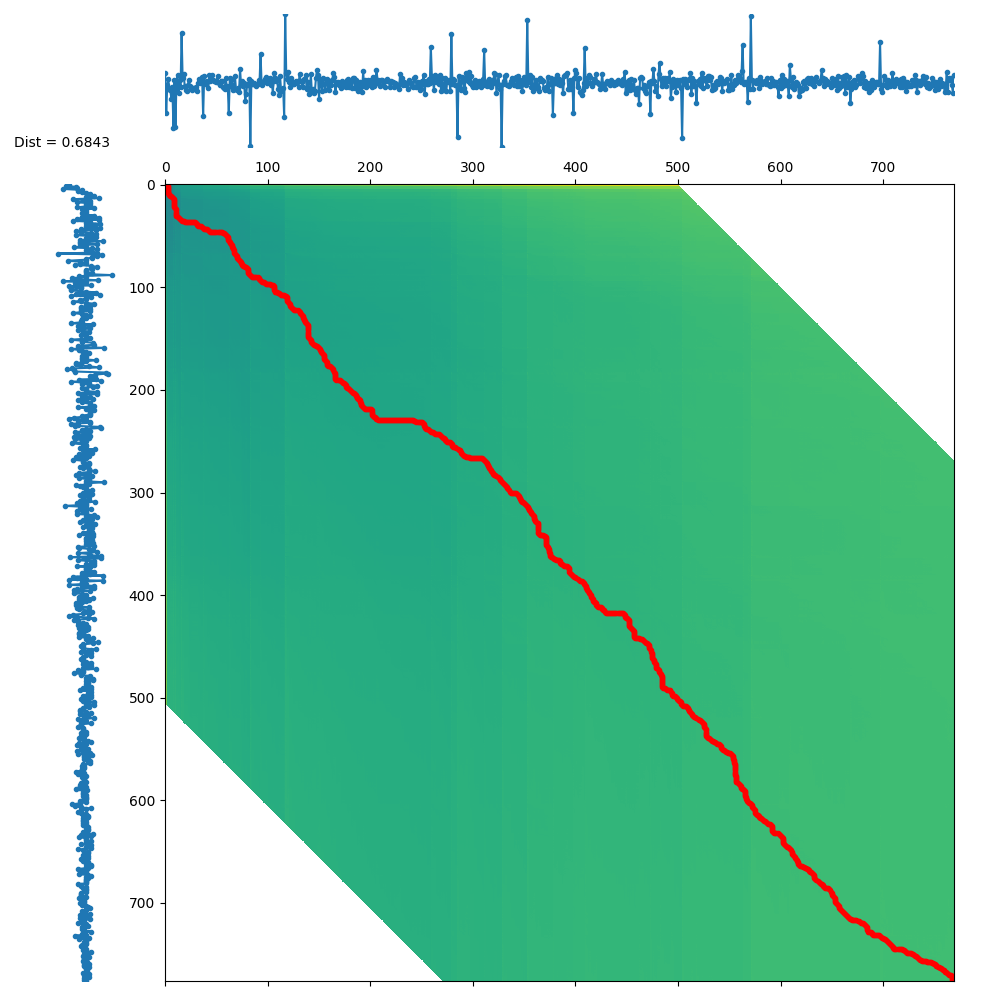}
    \caption{Dynamic Time Warping (DTW) distance}
    \label{fig:DTWD}
\end{figure}

Our Dynamic Time Warping (DTW) score (0.6843) shown in Figure~\ref{fig:DTWD} is lower than that reported in \cite{orlandi2024enhancing} (1.954), indicating that our QGAN-generated synthetic data more accurately capture the temporal dynamics of the historical data set. The experimental results demonstrate that the QGAN approach successfully generates synthetic time series data with high fidelity, effectively capturing the underlying dynamics of industrial bioprocesses.

A key measure of success is the Dynamic Time Warping (DTW) score, which evaluates the similarity between the generated and real time series data. The QGAN achieved a DTW score of 0.6843, which is lower than previously reported in \cite{orlandi2024enhancing} in financial data analysis, indicating improved temporal alignment and pattern preservation, while trained in data that are not exclusively presented as Gaussian, an interesting observation.

Dynamic Time Warping (DTW) represents a powerful technique for measuring similarity between temporal sequences that may exhibit varying lengths or temporal misalignments. In contrast to conventional Euclidean distance measures, DTW accommodates the stretching and compression of time axes to achieve optimal sequence alignment, even when patterns vary in speed or phase relationships. This characteristic makes DTW particularly valuable for comparing temporal patterns where key features may manifest at different time points across sequences.

The DTW algorithm operates by identifying an optimal warping path that minimizes the cumulative distance between corresponding elements of two time series while respecting monotonicity and boundary constraints. The warping path creates a non-linear mapping between sequence indices, allowing similar patterns to be matched even when they occur at different temporal positions or exhibit different durations. This alignment process ensures that DTW captures the underlying structural similarities between sequences regardless of local temporal variations.

The selection of DTW for temporal sequence comparison stems from its ability to accommodate local temporal distortions while preserving the overall sequential structure. Unlike rigid distance metrics, DTW provides robust similarity measurements for time-dependent data by dynamically adjusting for speed variations, length differences, and localized temporal shifts that commonly occur in real-world temporal patterns. The algorithm's flexibility in handling these temporal discrepancies makes it particularly suitable for applications where exact temporal correspondence cannot be assumed.

Our evaluation framework employs a multi-faceted approach to synthetic data validation, selecting metrics that collectively assess different aspects of data fidelity essential for bioprocess applications. DTW distance was chosen as our primary temporal similarity metric due to its robustness to time series alignment issues and its ability to capture non-linear temporal dependencies characteristic of bioprocess dynamics, where process variations can cause temporal shifts in key events such as growth phases or metabolite production peaks. The Wasserstein distance (Earth Mover's distance) provides a theoretically grounded measure of distributional similarity that is particularly sensitive to tail behaviors and outliers, critical considerations in bioprocess data where extreme values often represent important process states or anomalies. Quantile-Quantile (QQ) plots offer visual and statistical validation of normality assumptions underlying many bioprocess models, while Probability Density Function (PDF) and Cumulative Distribution Function (CDF) comparisons provide complementary perspectives on distributional fidelity, with PDFs revealing local density characteristics and CDFs offering global distributional alignment assessment.

To increase the potency of our evaluation framework, several additional metrics could be incorporated. The Maximum Mean Discrepancy (MMD) with radial basis function kernels would provide a non-parametric statistical test for distributional similarity with strong theoretical guarantees. Mutual Information estimation could quantify the preservation of feature dependencies crucial for multivariate bioprocess data, while the Kolmogorov-Smirnov test would offer formal statistical validation of distributional equivalence. For temporal aspects, metrics such as spectral density comparison through power spectral density analysis could validate frequency domain characteristics, and Granger causality tests could assess the preservation of causal relationships between process variables. Additionally, domain-specific bioprocess metrics such as correlation preservation in key process indicators (pH, dissolved oxygen, substrate concentration) and maintenance of process constraint relationships (mass balance equations, stoichiometric ratios) would provide application-specific validation. The integration of these complementary metrics would create a more complete validation framework, particularly important for safety-critical bioprocess applications where the quality of synthetic data directly impacts process optimization and monitoring system performance.

\section{Discussion}
\label{sec:Discussion}
This study presents a novel application of  QGANs for generating synthetic time-series data to address the pervasive data scarcity challenge in industrial bioprocess engineering. By leveraging PQCs within the generator, our framework captures complex temporal dependencies inherent in bioprocesses, resulting in synthetic data with demonstrably high fidelity, as evidenced by metrics such as DTW and distributional analyses.

The motivation for this work lies in a critical bottleneck of bioprocess development: real experimental data is expensive, time-consuming to generate, and often limited by regulatory and operational constraints. In addition, many key process variables, such as cell viability, product quality, or metabolic activity, are challenging to measure online, forcing the use of soft sensors or virtual inference models. However, building these predictive models requires large and diverse datasets. Synthetic data generation emerges as a necessary solution to enable accurate soft sensor development and improve process monitoring, optimization, and control.

Our results demonstrate that QGANs can produce synthetic multivariate time-series data that preserve both global statistical properties and local temporal dynamics of real processes. This capability is essential for training machine learning models that act as proxies for hard-to-measure quality indicators, ultimately aiding in maximizing product yields and ensuring process robustness. The ability of quantum circuits to efficiently represent complex probability distributions, thanks to properties like superposition, entanglement, and interference, offers a theoretical advantage in modeling non-linear, high-dimensional data typical of industrial bioprocesses.

Practical deployment of QGANs faces several challenges. Current quantum hardware limitations, including limited qubit counts, decoherence, and noise, restrict applicability to larger, more complex datasets. Furthermore, while our study focused on PQC-based generators, alternative quantum and hybrid classical-quantum architectures may offer additional benefits in generating diverse and realistic synthetic data. Exploring these architectures, optimizing circuit depth, and refining hybrid training strategies are promising future directions.

\begin{figure*}
    \centering
    \includegraphics[width=1\linewidth]{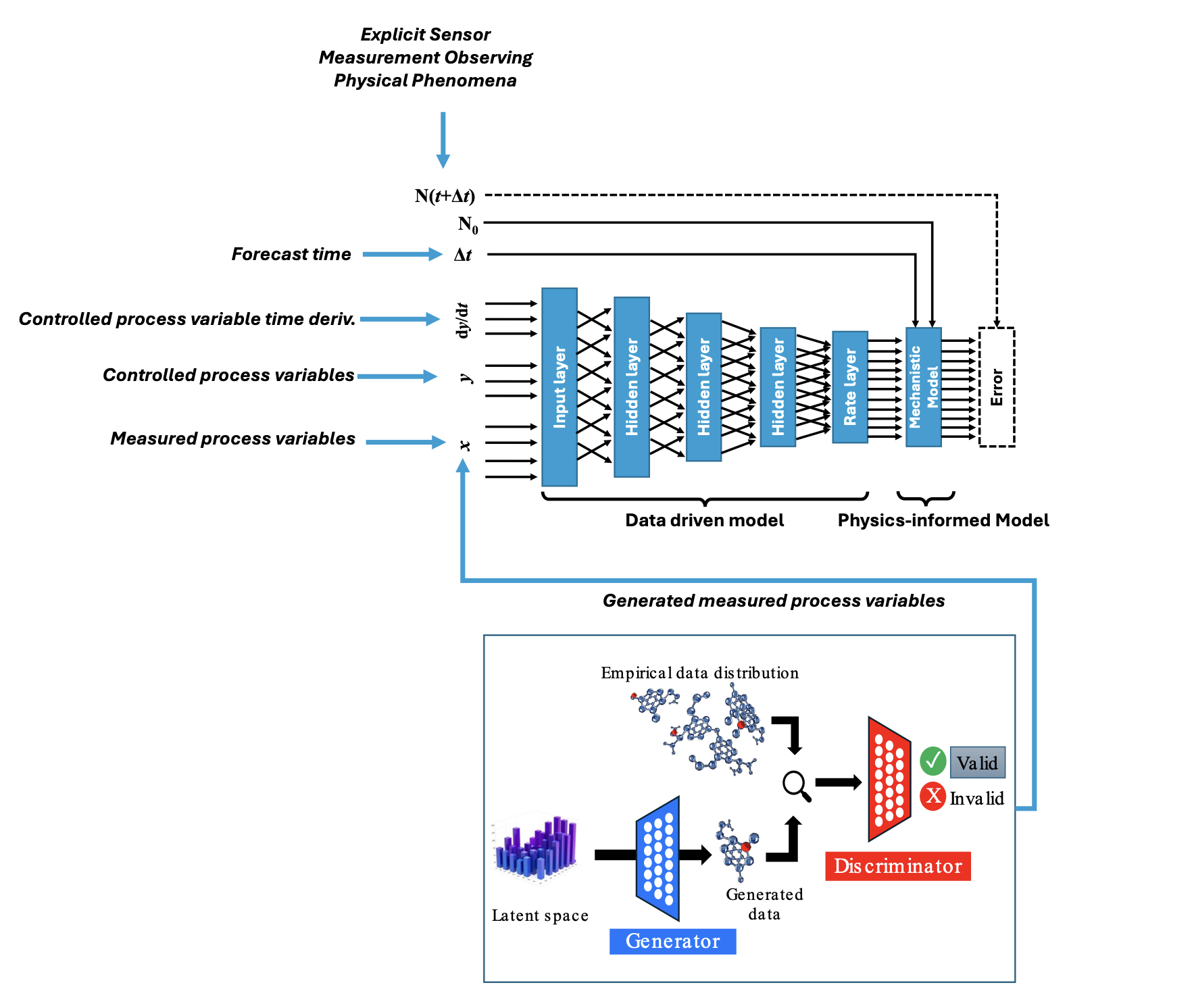}
    \caption{QGAN and synthetic data implementation within a hybrid-model approach to process monitoring}\label{fig:qGAN Hybrid}
\end{figure*}

The Hybrid-GAN-mechanistic structure represents an advanced architecture that integrates generative adversarial networks (GANs) with physics-informed mechanistic components to enhance predictive modeling in data-scarce environments such as the case stipulated in the current study. As can be observed in \ref{fig:qGAN Hybrid}, the quantum distributions obtained from the depth of a quantum hardware are used as a source or richer distributions for the generator. Although we propose using quantum distributions, it is important to note that we do not claim that a classical distribution, such as a Gaussian distribution \textit{cannot} be used in the above structure.

The core mathematical formulation of this hybrid structure is presented in Equation (9), which extends the standard GAN objective with a physics regularization term to ensure adherence to domain-specific constraints.

\begin{equation}
\begin{split}
    &E_{x \sim p_{data}}[\log D(x)] + E_{z \sim p_z}[\log(1-D(G(z)))] \\
    &\quad + \lambda E_{z \sim p_z}[dtd M(G(z)) - f(M(G(z)), t, \theta)^2]
\end{split}
\end{equation}

\begin{figure}
    \centering
    \includegraphics[width=1\linewidth]{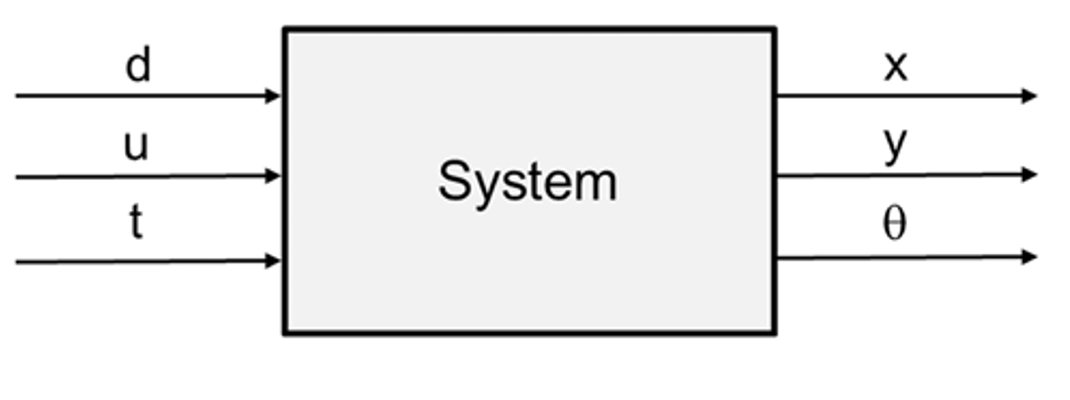}
    \caption{General representation of a (bio)chemical process using a black-box/data-driven model}\label{fig:gen_rep}
\end{figure}

\begin{figure}
    \centering
    \includegraphics[width=1\linewidth]{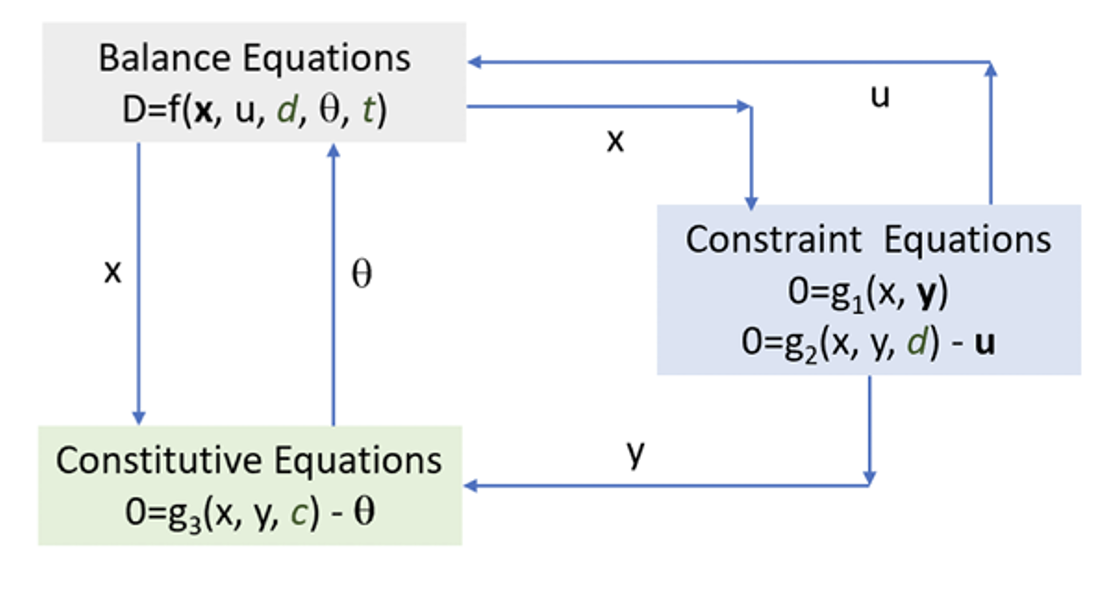}
    
    \caption{Mechanistic representation of the same process. Reprinted with permission from Elsevier \cite{mansouri2025models} }\href{applewebdata://63B383C9-234D-41D5-A9D6-D52B30698E25\#_msoanchor_1}{\label{fig:mech_rep}}
\end{figure}

In Equation (9) , $G$ denotes the generator mapping latent variables $z$ to generated process variables  $\hat{x}$, $D$ is the discriminator assessing authenticity; $M$ is the mechanistic operator projecting generated data onto physical states; $f$ encapsulates the governing dynamics (e.g., ordinary differential equations); $\lambda$ is a weighting hyperparameter for the physics loss, thus, balancing data fidelity and physical consistency; $\theta$ represents model parameters coming from the set of constitutive equations (e.g. reaction kinetics, particle size distribution, growth rates, etc); and $M(G(z))$ maps generated data to physical states. This optimization balances statistical fidelity from empirical data distributions ($p_{data}$) with physical consistency, minimizing residuals from time derivatives. In order to better describe $M$ for the case of a (bio)chemical system, see  Figure \ref{fig:gen_rep} and Figure \ref{fig:mech_rep}. Figure \ref{fig:gen_rep} shows the input-output structure of a black-box or data- based model. Variables in the input and output streams are measured for different conditions of process operation, for example, temperature and volume of a reactor, heat addition-removal in a heat exchanger, reflux rate in a distillation column, or rotating speed of a stirrer in a mixer, or aeration rate in a bioreactor, or light intensity in a photobioreactor. In Figure \ref{fig:mech_rep}, $c$ is a vector of model parameters, $d$ is a vector of specified variables, $t$ is an independent variable, $u$ is a vector of design variables, $x$ is a vector of process variables, $y$ is a vector of measured variables, and $\theta$ is a vector of physico-chemical properties as also described earlier.


This structure complements hybrid machine learning approaches prevalent in chemical and biological process modeling applications, particularly when data availability is limited (see Nielsen et al.,\cite{nielsen2020hybrid}  Nazemzadeh et al.,\cite{nazemzadeh2021integration}  and Ehtesham et al.,\cite{ehtesham2025dynamics} ) Hybrid modeling typically combines parametric models—derived from first-principles knowledge such as reaction kinetics or mass balances—with nonparametric, data-driven techniques like neural networks to mitigate the challenges of sparse datasets. \cite{sharma2022hybrid} For example, hybrid models augment mechanistic simulations with machine learning to improve accuracy in bioprocessing, where experimental data are often costly and scarce due to lack of process analytical technology (PAT) or regulatory restrictions (especially in biopharmaceutical production processes), enabling reliable predictions for process optimization and control. \cite{O'Brien2021A}  The proposed Hybrid-GAN as a future step for the current work, extends this paradigm by incorporating adversarial training, which generates synthetic data to augment limited empirical samples, for example Optical Density (OD) as discussed here. This is like physics-informed neural networks (PINNs) that embed governing equations into network architectures for solving ODEs. This connection underscores the Hybrid-GAN's utility in addressing small data challenges, where pure data-driven models fail due to insufficient training samples, by infusing prior scientific knowledge to produce robust, generalizable predictive models. Table \ref{tbl:various_approaches} lists a comparative overview of individual methods discussed in this work and the advantages of Hybrid-GAN structures which is the subject of future works based on the current study.

\begin{table}
\small
\caption{Comparison of various modeling approaches used in data-scarce environments, highlighting their strengths and limitations using check (\textbf{\ding{51}}) and cross (\textbf{\ding{55}}) marks.}
\label{tbl:various_approaches}
\resizebox{\columnwidth}{!}{%
\begin{tabular}{p{2cm}p{2cm}p{2cm}p{2cm}p{2cm}p{2cm}}
\hline
\textbf{Method} & \textbf{Data Efficiency} & \textbf{Physical Consistency} & \textbf{Generative Capability} & \textbf{Interpretability} & \textbf{Robust to Small Data}\\
\hline
Pure Data-Driven Models & \textbf{\ding{55}}& \textbf{\ding{55}}& \textbf{\ding{51}}& \textbf{\ding{55}}& \textbf{\ding{55}} \\
\midrule
Mechanistic Models & \textbf{\ding{51}}& \textbf{\ding{51}}& \textbf{\ding{55}}& \textbf{\ding{51}}& \textbf{\ding{51}}\\
\midrule
Hybrid Models (Mech + ML) & \textbf{\ding{51}}& \textbf{\ding{51}}& \textbf{\ding{55}}& \textbf{\ding{51}}& \textbf{\ding{51}}\\
\midrule
PINNs & \textbf{\ding{51}}& \textbf{\ding{51}}& \textbf{\ding{55}}& \textbf{\ding{51}}& \textbf{\ding{51}}\\
\midrule
Standard GANs & \textbf{\ding{55}}& \textbf{\ding{55}}& \textbf{\ding{51}}& \textbf{\ding{55}}& \textbf{\ding{55}}\\
\midrule
Hybrid-GAN (Proposed) & \textbf{\ding{51}}& \textbf{\ding{51}}& \textbf{\ding{51}}& \textbf{\ding{51}}& \textbf{\ding{51}} \\
\midrule
\end{tabular}}
%}
\end{table}

Physical sensor measurements, i.e. PAT, play a pivotal role in the Hybrid-GAN framework, serving as explicit observations ($y$) of underlying phenomena that ground the model in empirical reality. These measurements, often derived from sensors monitoring variables such as temperature, concentration, pH, or pressure in bio(chemical) reactors including the photobioreactor discussed in this work, provide direct validation points for the mechanistic component, \cite{hermann2025hybrid} By incorporating $y$ into the ODE formulation (e.g., $dy/dt = f(y, t, \theta) $ describing a process model), the model ensures that generated outputs $\hat{x}$ align with observable physical states, thereby constraining the latent space exploration and preventing deviations from feasible dynamics. This integration enhances model interpretability and reliability, as the physics-informed loss in Equation (8) penalizes inconsistencies between simulated and measured derivatives.

Regarding overfitting and underfitting, the Hybrid-GAN mitigates these issues through its dual regularization mechanism. Overfitting, common in data-driven models trained on limited noisy data, is addressed by the physics term in Equation (9), i.e. $M(G(z))$, which imposes hard constraints from mechanistic laws, reducing the model's tendency to memorize artifacts and promoting generalization. Underfitting, where models fail to capture complex patterns, is countered by adversarial training, which enriches the data distribution via synthetic samples, while the mechanistic operator ensures essential dynamics are not overlooked \cite{rice2020overfittingadversariallyrobustdeep} . Thus, the framework fosters models that are both data-efficient and physically sound.

\section*{Conclusions}

This work presents a novel application of Quantum Wasserstein Generative Adversarial Networks with Gradient Penalty (QWGAN-GP) for synthetic time-series data generation in industrial bioprocess monitoring, addressing the critical challenge of data scarcity that has long hindered the development of robust predictive models and soft sensors in bioprocess engineering.

\subsection{Key Contributions and Findings}

Our research demonstrates four principal contributions to the field. First, we introduced a comprehensive eight-stage, decision-driven pipeline that seamlessly integrates sensor feasibility assessment, mechanistic modeling, data-driven learning, and quantum synthetic data generation within a unified feedback loop. This framework provides a systematic approach to bioprocess monitoring that can adapt to varying data availability constraints and sensor limitations.

Second, we successfully developed and validated a QWGAN architecture specifically optimized for bioprocess applications, where the generator component employs a Parameterized Quantum Circuit (PQC) with three layers of quantum gates, including Hadamard gates for superposition creation, rotation gates for amplitude control, and CNOT gates for entanglement generation. This quantum-enhanced generator demonstrated superior performance in capturing the complex temporal dependencies and non-linear dynamics characteristic of biological systems.

Third, our empirical validation on real-world photobioreactor cultivation data confirms the effectiveness of the proposed approach. The QWGAN achieved a Dynamic Time Warping (DTW) score of 0.6843, representing improved temporal alignment compared to previously reported methods. The synthetic data preserved global statistical properties, as evidenced by strong normality alignment in quantile-quantile analyses, faithful reproduction of auto-correlation structures, and accurate preservation of probability density and cumulative distribution functions.

Fourth, we have contributed to open science practices by making our code and datasets publicly available, facilitating independent verification and accelerating future advances in quantum-enhanced bioprocess engineering.

\subsection{Theoretical and Practical Implications}

The theoretical significance of this work lies in demonstrating how quantum computational advantages can be effectively applied to address real-world engineering challenges. Our results suggest that quantum circuits can efficiently represent complex multivariate probability distributions typical of bioprocess data, potentially offering computational advantages over classical approaches when working with limited datasets.

From a practical perspective, this research addresses a fundamental bottleneck in bioprocess development where experimental data generation is constrained by high operational costs, regulatory requirements, and time limitations. By generating high-fidelity synthetic data that preserves the essential characteristics of real bioprocess dynamics, our approach enables the development of more robust soft sensors for monitoring critical quality indicator variables such as biomass concentration, product quality, and metabolic activity—variables that are typically difficult or expensive to measure online.

\subsection{Future Research Directions}

While our results are promising, several avenues for future investigation warrant attention. The scalability of quantum circuits remains constrained by current NISQ device limitations, including limited qubit counts, decoherence effects, and circuit depth restrictions. Future work should explore the integration of error mitigation techniques and investigate the performance scaling with larger, more complex bioprocess datasets as quantum hardware capabilities advance.

The proposed Hybrid-GAN-mechanistic architecture represents a particularly promising direction, where physics-informed constraints from first-principles models are integrated with adversarial training to ensure both statistical fidelity and physical consistency. This approach could address limitations in current synthetic data generation methods by incorporating domain knowledge constraints while maintaining the flexibility of data-driven approaches.

Additionally, extending the framework to multivariate, multimodal bioprocess datasets and exploring alternative quantum generative models such as quantum diffusion models or quantum variational autoencoders could further enhance the versatility and performance of synthetic data generation for bioprocess applications.

\subsection{Concluding Remarks}

This research establishes quantum-enhanced generative adversarial networks as a viable and potentially superior approach to addressing data scarcity challenges in industrial bioprocess engineering. By successfully demonstrating that QWGANs can generate synthetic time-series data while preserving the complex statistical properties of real bioprocess systems, we have opened new possibilities for advanced bioprocess monitoring, optimization, and control strategies.

The convergence of quantum computing and bioprocess engineering represents an emerging frontier with significant potential for industrial impact. As quantum hardware continues to mature and become more accessible, the methodologies developed in this work provide a foundation for next-generation bioprocess management systems that can operate effectively under data-constrained conditions while maintaining the precision and reliability required for industrial applications.

Our findings suggest that the integration of quantum machine learning techniques with traditional bioprocess engineering approaches offers a promising pathway toward more efficient, robust, and economically viable biomanufacturing processes, ultimately contributing to advances in pharmaceutical production, sustainable materials development, and biotechnology applications where data limitations have historically impeded progress.

\section*{Author contributions}

\textbf{Shawn Gibford:} Conceptualization, Methodology, Software, Investigation, Writing - Original Draft.  \textbf{Mohammed Reza Boskabadi:} Conceptualization,  Writing - Review \& Editing. \textbf{Christopher J. Savoie:} Conceptualization, Supervision, Writing - Review \& Editing. \textbf{Seyed Soheil Mansouri} Conceptualization, Supervision, Writing - Review \& Editing, Project administration, Funding acquisition.

\section*{Conflicts of interest}
There are no conflicts to declare.

\section*{Data availability}

The complete code implementation for the Quantum Wasserstein Generative Adversarial Network with Gradient Penalty (QWGAN-GP) and experimental bioprocess datasets supporting the results of this article are freely available at https://github.com/shawngibford/qgan. The repository includes the parameterized quantum circuit implementation, classical discriminator architecture, data, evaluation metrics, and reproducibility notebooks. The photobioreactor cultivation time-series data, including optical density measurements from the LUCY® system, are provided in standardized CSV format. All code is implemented in Python using PennyLane for quantum circuit simulation and PyTorch for classical neural network components.

\section*{Acknowledgments}

The authors would like to thank and acknowledge several individuals that took part in this research and helped make it a success. Alireza Mehrdadfar for operation of the Lucy bioreactor and subsequent collection and organization of the data it produced, and his assistance with producing some of the figures seen in this article.

%%%END OF MAIN TEXT%%%

%%%BALANCE COLUMNS ON FINAL PAGE%%%
\balance

%%%BIBLIOGRAPHY%%%
\bibliography{bib}
\bibliographystyle{abbrv}
\end{document}